\documentclass[12pt, preprint]{aastex}

\def\deg{\ifmmode^\circ\else$^\circ$\fi}

\def\mic{~$\mu$m}

\def\mic{$\mu${\rm m}}

\def\arcs{\ifmmode {''}\else $''$\fi}
\def\arcm{\ifmmode {'}\else $'$\fi}
\def\parcs{\sa=.07em \sb=.03em
     \ifmmode $\rlap{.}$^{\scriptscriptstyle\prime\kern -\sb\prime}$\kern -\sa$
     \else \rlap{.}$^{\scriptscriptstyle\prime\kern -\sb\prime}$\kern -\sa\fi}
\def\parcm{\sa=.08em \sb=.03em
     \ifmmode $\rlap{.}\kern\sa$^{\scriptscriptstyle\prime}$\kern-\sb$
     \else \rlap{.}\kern\sa$^{\scriptscriptstyle\prime}$\kern-\sb\fi}

\def\spose#1{\hbox to 0pt{#1\hss}}
\def\simlt{\mathrel{\spose{\lower 3pt\hbox{$\mathchar"218$}}
     \raise 2.0pt\hbox{$\mathchar"13C$}}}
\def\simgt{\mathrel{\spose{\lower 3pt\hbox{$\mathchar"218$}}
     \raise 2.0pt\hbox{$\mathchar"13E$}}}
\def\lsim{\rlap{$<$}{\lower 1.0ex\hbox{$\sim$}}}
\def\gsim{\rlap{$>$}{\lower 1.0ex\hbox{$\sim$}}}

\begin{document}

\title{The Far- and Mid-Infrared/Radio Correlations in the {\it Spitzer} Extragalactic First Look Survey}

\author{P. N. Appleton\altaffilmark{1}, D. T. Fadda\altaffilmark{1},
F. R. Marleau\altaffilmark{1}, D. T. Frayer\altaffilmark{1},
G. Helou\altaffilmark{1}, J. J. Condon\altaffilmark{2},
P. I. Choi\altaffilmark{1}, L. Yan\altaffilmark{1},
M. Lacy\altaffilmark{1}, G. Wilson\altaffilmark{1},
L. Armus\altaffilmark{1},  S. C. Chapman\altaffilmark{4},
F. Fang\altaffilmark{1}, I. Heinrichson\altaffilmark{1},
M. Im\altaffilmark{3}, B. T. Jannuzi\altaffilmark{5},
L.J. Storrie-Lombardi\altaffilmark{1}, 
D. Shupe\altaffilmark{1}, B.T. Soifer\altaffilmark{1},
G. Squires\altaffilmark{1}, H. I. Teplitz\altaffilmark{1}}

\altaffiltext{1}{{\it Spitzer} Science Center, MS 220-6, Caltech, Pasadena, CA 91125. apple@ipac.caltech.edu}
\altaffiltext{2}{NRAO, 520 Edgemont Road, Charlottesville, VA 22903-2475}
\altaffiltext{3}{Seoul National University, Shillim-dong, San 56-1, Kwanak-Gu,
Seoul, S. Korea}
\altaffiltext{4}{MS 320-47,Caltech, Pasadena, CA 91125.}
\altaffiltext{5}{NOAO, 950 North Cherry Avenue, P.O. Box 26732, Tucson, Arizona 85726}
\begin{abstract}
  Using the {\it Spitzer Space Telescope} and the VLA, we present the
  first {\it direct} evidence that the well-known far-infrared/radio
  correlation is valid to cosmologically significant redshift. We also
  confirm, with improved statistics compared with previous surveys, a
  similar result for the Mid-IR/radio correlation. We explore the
  dependence of monochromatic $q_{24}$ and $q_{70}$ on $z$.  The
  results were obtained by matching {\it Spitzer} sources at 24 and
  70\mic ~with VLA 1.4GHz $\mu$Jy radio sources obtained for the {\it
  Spitzer} FLS. Spectroscopic redshifts have been obtained for over
  500 matched IR/radio sources using observations at WIYN, Keck and
  archival SDSS data extending out to $z~>$~2. We find that $q_{24}$
  shows significantly more dispersion than $q_{70}$. By comparing the
  observed fluxes at 70, 24 and 4.5\mic ~with a library of SED
  templates, we find that the larger dispersion in $q_{24}$ is
  predictable in terms of systematic variations in SED shape
  throughout the population.  Although the models are not able to
  encompass the full range of observed behavior (both the presence of
  either extremely flat or extremely steep IR SEDs), the fitting
  parameters were used to ``k-correct'' the higher-z galaxies which
  resulted in a reduced scatter in $q$. For comparison, we also
  corrected these data using the SED for M82.  The results for 24 and
  70\mic ~provide strong consistent evidence for the universality of
  the mid-IR/radio and far-IR/radio correlations out to redshifts of
  at least $z$~=~1.
\end{abstract}


\section{Introduction}
It has now been more than a quarter of a century since the first hints
emerged of a very tight correlation between the global far--infrared
(FIR) and radio emission from galaxies. Initially based on comparisons
between challenging ground-based 10{\mic}/radio correlations of 
small samples of galaxies (van der Kruit 1973; Condon et al. 1982;
Rikard \& Harvey 1984), the remarkable nature of the FIR/radio
correlation became increasingly apparent with larger IR--samples
obtained from the Infrared Astronomical Satellite (IRAS) all sky
survey (Dickey \& Salpeter 1984 ; De Jong et al. 1985; Helou, Soifer,
\& Rowan-Robinson 1985; Condon \& Broderick 1986; Wunderlich et
al. 1987; Hummel et al.  1988; Fitt, Alexander, \& Cox 1988).  IRAS
showed that the FIR-radio correlation appeared to apply 
to a wide range of Hubble-types, extending over more than three
orders of magnitude in IR luminosity from dust-rich dwarfs to
ultra-luminous infrared galaxies (ULIRGS). 

Though revolutionary, IRAS could only systematically study the
relatively nearby universe. The Infrared Space Observatory (ISO)
probed, for the first time, galaxy populations in the higher-z
universe, but had limited long-wavelength coverage and sensitivity.
Based on the deepest ISO fields, it was evident that even at 15\mic ~
there was a loose correlation between the Mid-IR (MIR) emission, and
the radio continuum (Cohen et al. 2000, Elbaz et al. 2002, Gruppioni
et al. 2003). Using M82 as a template, Garrett (2002) extrapolated
from 15\mic ~to longer wavelengths, and suggested that the
Far-IR/radio correlation probably extends to $z >$ 1. At the other
extreme, sub-mm observations using instruments like SCUBA, began to
reveal a new population of likely star-formation active
galaxies. Although the statistics and redshift determinations are
still growing (see Chapman et al. 2003), the correlation between
$\mu$Jy radio and SCUBA sources suggest that the FIR/radio correlation
is likely to be valid at much higher redshift (see Carilli \& Yun
1999, Ivison et al. 1998, Frayer et al. 1999, Ivinson et
al. 2002). The universality of the correlation, if confirmed, is
important because it implies that galaxies observed at large look-back
times share many of the same properties as fully-formed mature
galaxies seen locally.

\section{Spitzer, VLA Radio and Spectroscopy Observations}

{\it Spitzer} observations\footnote{This work is based [in part] on
observations made with the Spitzer Space Telescope, which is operated
by the Jet Propulsion Laboratory, California Institute of Technology
under NASA contract 1407.}  of the First Look Survey (FLS) were
obtained from 1--11 December 2003 (Storrie-Lombardi et al. in prep.,
Fadda et al. 2004). A total of 4 sq. degree was covered by both the
IRAC (3.6, 4.5, 5.8, 8.0\mic; Fazio et al. 2004--this volume) and MIPS
(24, 70 and 160\mic; Rieke et al. 2004--this volume) instruments at
medium depth (5 $\times$ 12s IRAC, 2 $\times$ 40s for MIPS 24, 1
$\times$40s 70\mic, and 2$\times$4s at 160\mic), and a smaller
``verification strip'' (0.25 square degree) was observed to an even
greater depth (4 $\times$ 40s). The IRAC and MIPS 24\mic ~data were
processed through standard SSC pipelines (Lacy, Fadda \& Frayer, in
prep.). The average FWHM of the point-spread-function of the final
maps were $\sim$2$^{\prime\prime}$ for the IRAC bands, and 6, 18 and
40$^{\prime\prime}$ for MIPS 24, 70\mic ~and 160\mic ~respectively.

IRAC sources were extracted (Lacy et al. in preparation) using the
package SExtractor (Bertin \& Arnouts 1996). MIPS~24 and MIPS~70\mic ~
sources were extracted from the images using the StarFinder method
(Diolaiti et al. 2000).  The formal 5-$\sigma$ sensitivities
are measured to be 6, 7, 45 and 32$\mu$Jy for IRAC (3.6, 4.5, 5.8 \&
8\mic) and 0.3, 30 and 100 mJy for MIPS (24,70 and 160\mic ~
respectively). Tests performed on data from both the full survey and
the verification strip at 24\mic ~and 70\mic ~indicate that the source
flux densities show good reliability and repeatability down to a flux
density of $\sim$0.5~mJy and 30~mJy respectively.

Radio observations were carried out by Condon et al.(2003) down to a
formal 5-$\sigma$ depth of 115 mJy with 35-overlapping B-array VLA
pointings (FWHM of restored synthesized beam = 5$^{\prime\prime}$),
revealing over 3500~$\mu$Jy radio sources.  This conservative limit is
appropriate for the entire survey, including the edges of the field
which were not surveyed as deeply as the inner regions. Given the
coverage of the MIPS and IRAC fields, which did not extend to the
outer edges of the VLA survey, we used a slightly deeper catalog
(S(20cm)$>$ 90$\mu$Jy prepared by JJC) more appropriate for
cross-matching with the {\it Spitzer} data. The deeper VLA catalog was
found to provide a high degree of reliability when cross-correlated
with deep R-band imaging of the same field.

Our sample consists of VLA radio sources with MIPS~24 and/or MIPS~70
counterparts and known spectroscopic redshifts.  Source matching
between the {\it Spitzer} and radio catalogs involved an automated
catalog search, followed by a visual inspection of the IR and radio
images (this was important because of the possible presence of noise
artifacts in the 70\mic ~images near the edges of scan legs). For the
IRAC and MIPS 24\mic ~source catalogs, we searched for IR counterparts
to VLA sources within a radius of r $<$ 2.5$^{\prime\prime}$ of the
VLA centroid. For the MIPS 70\mic ~images, the search was performed
out to a radial distance of 12$^{\prime\prime}$ because of the larger
effective beam-size of the 70\mic ~observations. For MIPS 24\mic ~and
70\mic ~, 508 and 227, matched sources were found respectively. Of the
$>$ 3000 matched IRAC/VLA sources, 412 4.5\mic ~sources were in common
with the MIPS/VLA sample used in this paper\footnote{The reason we
used IRAC 4.5\mic ~sources in this paper is because we have strong
evidence in our sample that the 5.8 and 8.0\mic~ sources are strongly
evolving in flux in the $z$ range 0 to 0.8: presumably due to the
shifting out of these bands of the 3.6, 6.2 and 7.7\mic (PAH) infrared
features. Using the 4.5\mic ~sources also increased the number of
MIPS/VLA matches compared with the less sensitive IRAC filters at
longer wavelengths}.  So few 160\mic ~detections were made that we
restricted ourselves to the shorter wavelengths.

Spectroscopy of likely MIPS sources was performed before the launch of
{\it Spitzer} by targeting VLA radio sources in two separate
surveys. The first comprised of $\sim$ 1200 radio sources, and were
observed using the NOAO-WIYN/Hydra fiber-system (Marleau et al. in
preparation) while the second survey yielded 80 radio sources targeted
with the KeckII/Deimos spectrograph. The latter was part of a larger
spectroscopy study of the FLS (Choi et al. in preparation). These data
were supplemented with archival redshifts from the Sloan Digital Sky
Survey.

\section{Model-dependent ``$k$-corrections''} 

In this paper we present primarily flux-density ratios between the IR
and the radio observations. Neglecting the $(1+z)^{-1}$ bandwidth
compression term of the frequency units of flux density (Weedman 1986,
Hogg et al. 2002), which cancels for flux ratios, the radio flux densities
were boosted by a factor $(1+z)^{0.7}$, where we assumed a synchrotron
power law of the form S $\propto {\nu}^{-0.7}$, a spectral index =
+0.7 (a value typical of an average steep spectrum radio source). The
boosting corrects the observed flux density at the emitted frequency
to the value it would have at the observed frequency in its rest
frame.

We approach the IR $k$-correction using two methods. Where 4.5, 24 and
70\mic ~ data were available, we compared these data with a set of
redshifted spectral energy distributions (SEDs) from Dale \& Helou
(2001; hereafter DH) and used these to compute the correction.  As we
shall see, the models only partly encompass the observed galaxy
behavior. Therefore, as a second approach (and to extend the SED
corrections to cover a larger number of 24\mic ~sources), we also
use a complete SED for M82 (Fadda et al. 2002) appropriately convolved
to the resolution of the Spitzer filters to correct these data.

The DH SEDs are a set of galaxy SEDs ranging from optical to sub-mm
wavelengths and covering 64 different levels of excitation, as
measured by the dust excitation parameter ${\alpha}$ ranging from
0.063 to 4.\footnote{The galaxy SEDare created by combining
individual' ``model'' SEDs according to a power-law distribution in 
$UdM(U)~\propto~U^{-\alpha}dU$, where $M(U)$ represents the dust mass
heated by radiation of energy density U.}.  As discussed in detail by
DH, values of $\alpha$ less than 1 correspond to a luminous starburst,
while larger $\alpha$ values are more appropriate for describing
quiescent systems.  To select the best suited template SED, we
computed the 4.5, 24 and 70\mic ~fluxes of each template redshifted to
the value appropriate for each galaxy.  We then use the observed 4.5,
24 and 70\mic ~flux densities (or 4.5 and 24\mic ~if 70\mic ~is
unavailable) to perform a ${\chi}^2$ fit to the {\it Spitzer} data.
We selected the model which returned the lowest value of
${\chi}^2$. The flux density at the observed frequency in the
best-fitting model SED was then compared with the same frequency in
the $z$ = 0 model to derive the $k$-correction for that target galaxy.

\section{Far and Mid-IR/Radio correlation as a Function of Redshift}

In Fig. 1a we show the monochromatic 20cm--versus--70\mic ~luminosity
correlation\footnote{Here the emitted luminosity is L$_{{\nu}e}$=
4$\pi$ d$_l$$^2$ S$_{{\nu}obs}$/(1+z) (Hogg et al. 2002), where
S$_{{\nu}obs}$ has already been k-corrected for an assumed SED as
previously discussed. The luminosity distance, d$_l$, assumes $H_0 =
70$ km s$^{-1}$\ Mpc$^{-1}$, $\Omega_M=0.3, \Omega_{\Lambda}=0.7$.}.
The plot shows that galaxies span four orders of magnitude in IR
luminosity, and that a small number of radio-dominant AGNs are seen
above the main trend.  However, because of the distance-squared
stretching of points in this diagram, we prefer to discuss the
correlation and its dependence on redshift in term of the more basic
observable parameters--namely $q_{ir}$, where $q_{ir}$ =
log($S_{ir}$/S$_{20cm}$), and $S_{ir}$ is the flux density in the
24\mic ~or 70\mic ~bands respectively. These monochromatic values for
$q$ represent the slope of the IR/radio correlation, and the
dispersion about the mean $q$ is indicative of the strength of the
correlation.\footnote{Ideally, it would be better to measure a
bolometric $q$, but insufficient data is available at longer
wavelengths to do this reliably (see Papovich \& Bell 2002). In
addition, monochromatic q relationships may be of great interest for
studies involving only one or two MIPS bands.}

Fig.~1b and c show the distribution of $q_{70}$ points before and
after the application of a $k$-correction using the SED
model-fitting method. Ignoring the points below $q$ = 1.6 which are
associated with radio-loud AGN, the figure shows a tight correlation
between radio and 70\mic ~emission over a wide range of redshifts
extending to $z \ge $~1.

\begin{figure}[t*]
\caption{\label{fig:1} a) The 20cm radio and 70\mic ~IR luminosity
correlation (see text) for the FLS long-wavelength sample, b) The
distribution of monochromatic $q_{70}$ values with redshift,
uncorrected and, c) k-corrected using the DH SED-fitting method
described in the text for the IR flux densities and assuming a
(1+z)$^{0.7}$ k-corrected (boosting) of the 1.4 GHz values.}
\end{figure}

We estimated the median and dispersion of the correlation using a
biweight-estimator discussed by Beers, Flynn \& Gebhardt (1990). This
method is resistant to outliers and robust for a broad range of
non-Gaussian underlying populations.  The central location (mean)
value and scale (dispersion) in $q_{70}$ before (2.16$\pm$0.17) and
after (2.15$\pm$0.16) correction appears to be invariant with $z$ up
to 1. The slight increase in $q_{70}$ above $z$ = 1 is probably the
result of small number statistics, and the strong Malmquist bias which
leads to very luminous galaxies appearing in the sample at these
redshifts. The highest redshift galaxies have luminosities of
comparible with that of ULRGS.

We show in Fig.~2a and b the corresponding $q$ diagrams for the 24\mic
~band.  The filled circles indicate galaxies that have a measured flux
at 70\mic. The dispersion in the $q_{24}$-values at zero redshift
(mean = 0.84$\pm$0.28) is a factor of 1.6 $\times$ larger than that
seen at 70\mic. A general flaring of the $q_{24}$ distribution with
$z$ is noticeable in the uncorrected data. Fig.~2b, which has been
corrected using the SED-fitting method, has fewer points than Fig.~2a
because the method relies on having detections at 70\mic. Since the
flaring is reduced after the correction, at least part of the higher-z
dispersion can be attributed to variations in SED, as reflected in the
statistics (mean = 0.94$\pm$0.23). The deviations from a constant q
for the highest-z galaxies may be due to Malquist bias which would
favor very luminousity ULIRGs which may show free-free absorption at
radio wavelengths. Fig.~2c contrasts the more definitive SED-fitting
method with the M82-correction method applied to the complete set of
24\mic ~detected galaxies. The formal scatter is also slightly reduced
(mean = 1.00$\pm$0.27), especially in the core of the distribution,
although outlying points are less well corrected, presumably because
M82 is not a good fit to all the SEDs (formally the M82 method reduces
the scatter as well as the SED method if the smaller number of points
only are considered, with the mean = 1.07$\pm$0.21).

\begin{figure}[t*]
\caption{\label{fig:2} $q_{24}$ as a function of redshift for a)
uncorrected and b) k-corrected using the SED-fitting method, and c)
correcting with an M82 template (see text).}
\end{figure}
 
The poor 24\mic ~correlations at low $z~<$~0.2
must reflect intrinsic variations in the radio/IR ratio within the
population. Insight into this can be seen in Fig.~3, which shows
$q_{24}$ as a function of the 24/70\mic ~color. This 
shows that, for a given color, the galaxies have a much smaller
$q$-dispersion. For example, it is clear that the majority of the
outlying points in the $q$-distribution are at the extremes of the color
distribution, either having very low ($<$0.04 S$_{24}$/S$_{70}$) or
very high ($>$ 0.2) flux ratios.  Note that the AGNs do not
follow the clear trend with color, allowing easy separation of this
population from the star forming systems.

Also plotted on Fig.~3 are the loci of model $\alpha$-parameter from
the SED library for three different redshift ranges. The solid lines
represent the model predictions for the full range of possible
$\alpha$-parameters represented in the model. The figure shows that
the general slope and behavior of the $q$-versus-color diagram is
extremely well represented, but the range of possible color-space is
too limited at present. It will be noted that many galaxies lie just
outside the range of coverage of the models at low-z, although at
high-z the models encompass more of the distribution. The consequence
of the incomplete coverage of the model parameters provides an
explanation for the apparent increase in $q$ with $z$ seen in
Fig.~2b. The model-results are promising, though presently limited in
scope. However, they do suggest that great improvements in the method
will come by the inclusion of more steeply falling SEDs in the
template library.

\section{Conclusions}

Observations with the {\it Spitzer Space Telescope} show that the
slope of the 70\mic ~FIR/radio correlation is constant to out to $z$ =
1, and that the dispersion about the mean is invariant to $z$ =
0.5. Further, using the higher sensitivity gained by observations at
24\mic ~, we show that, despite the larger scatter, the Mid-IR/radio
correlation has constant slope out to $z$ = 2. In a lambda-dominated
universe, this implies that the ``conspiracy'' of finely tuned factors
which have created the correlation in nearby galaxies existed in the
potentially more primitive galaxies which we observe more than
8-10~Gyrs in the past.  Unlike some dwarf galaxies seen locally, which
are unable to retain their radio-emitting plasma (and deviate from the
FIR-radio correlation--Klein et al. 1984, Chi \& Wolfendale, 1990),
these distant galaxies must be able to contain their radio-emitting
plasma long-enough to fall on the FIR/radio relationship. Future work
will concentrate on understanding the systematic effects which lead to
the increased spread in the strength of the MIR correlation, thus
allowing the technique to be applied to much deeper surveys where
significant overlap with powerful sub-mm galaxy population will be possible.

\begin{figure}[t*]
\caption{\label{fig:3} The trend of $q_{24}$ with mid-far IR
color. Open circles = galaxies with $z~<$~0.5, and filled circles show
galaxies with $z~>+$~0.5. The solid lines show the range of
q-parameter and colors predicted by the DH model SEDs for increasing
values of $\alpha$-parameter. The range is from $\alpha$ = 0.063 to 4
(ending with arrow). Curves for three redshift ranges are
represented. }
\end{figure}

\references

\reference{}Bertin, E. \& Arnouts, S. 1996, A\&AS, 117, 393
\reference{}Carilli, C. L. \& Yun, M. S. 1999, ApJL, 513, L13  
\reference{}Chapman, S. C., Blain, A. W., Ivison, R. J. \& Smail, I. R. 2003, Nature, 422, 695 
\reference{}Chi, X. \& Wolfendale, A. W. 1990, MNRAS, 245, 101
\reference{}Cohen, J. et al. 2000, ApJ, 538, 29 
\reference{}Condon, J. J., Cotton, W. D., Yin, Q. F., Shupe, D. L.,
 Storrie-Lombardi, L. J., Helou, G., Soifer, B. T., 
 Werner, M. W. 2003, AJ, 125, 2411
\reference{}Condon, J. J., Condon, M. A., Gisler, G. \& Puschell,
J. 1982. ApJ, 252,102
\reference{}Condon, J. J. \& Broderick, J. J. 1986, AJ, 92, 94

\reference{}Dale, D. A. \& Helou, G., Contursi, A. , Silbermann, N. A. \& Kolhatkar, S. 2001, ApJ, 549, 215  [DH]
\reference{}de Jong, T., Klein, U., Wielebinski, R. \& Wunderlich, E. 1985, A\&A, 147, L6
\reference{}Diolaiti, E, Bendinelli, O., Bonaccini, D., Close, L., Currie, D., Parmeggiani, G, 2000, A\&AS, 147, 335
\reference{}Dickey, J. M., Salpeter, E. E. 1984, ApJ, 284, 461
\reference{}Elbaz, D., Cesarsky, C. J., Chanial, P., Aussel, H., Franceschini, A., Fadda, D. \& Chary, R. R. 2002, A\&A, 384, 848
\reference{}Fadda D. et al. 2002, A\&A, 383,838
\reference{}Fadda, D., Jannuzi, B., Ford, A., \& Storrie-Lombardi, L.J. 2004, AJ, in press [astro-ph/0403490]
\reference{}Frayer, D. T., Ivison, R. J., Scoville, N. Z., 
 Evans, A. S., Yun, M. S., Smail, I. Barger, A. J., Blain, A. W., Kneib, J.-P. 1999 ApJL, 514, L13
\reference{}Fitt, A. J., Alexander, P. \& Cox, M. J. 1988. MNRAS, 233, 907
\reference{}Garrett, M. 2002, A\&A, 384, L19
\reference{}Gruppioni et al. 2003, MNRAS, 341, L1
\reference{}Helou, G., Soifer, B. T. \& Rowan-Robinson, M. 1985. ApJL, 298, L7 
\reference{} Hogg, D. W., Baldry, I. K., Blanton, M. R. \& Eisenstein, D. J. 2002, astro-ph-0210394
\reference{}Hummel, E., Davies, R. D. \& Wolstencroft, R. D., van der Hulst, J. M., Pedlar, A. 1988. A\&A, 199, 91
\reference{}Ivison, R. J., Smail, I., Le Borgne, J.-F., Blain, A. W., \& Kneib, J.-P., et al. 1998, MNRAS, 298, 583
\reference{}Ivison, R. J. et al. 2002, MNRAS, 337, 1
\reference{}de Jong, T., Klein, U., Wielebinski, R. \& Wunderlich, E. 1985. AA, 147, L6
\reference{}Klein, U., Wielebinski, R., Thuan, T. X. 1984, A\&A, 141, 241
\reference{}Papovich, C. \& Bell, E. F. 2002, ApJL, 579, L1  
\reference{}Rickard, L. J., Harvey, P. M. 1984 AJ, 89, 1520
\reference{}van der Kruit, P.C. 1973, A\&A, 29, 263 
\reference{}Weedman, D. W. 1986, ``Quasar Astronomy'', (CUP, Cambridge), p61
\reference{}Wunderlich, E., Klein, U. \& Wielebinski, R. 1987. A\&AS 69, 487

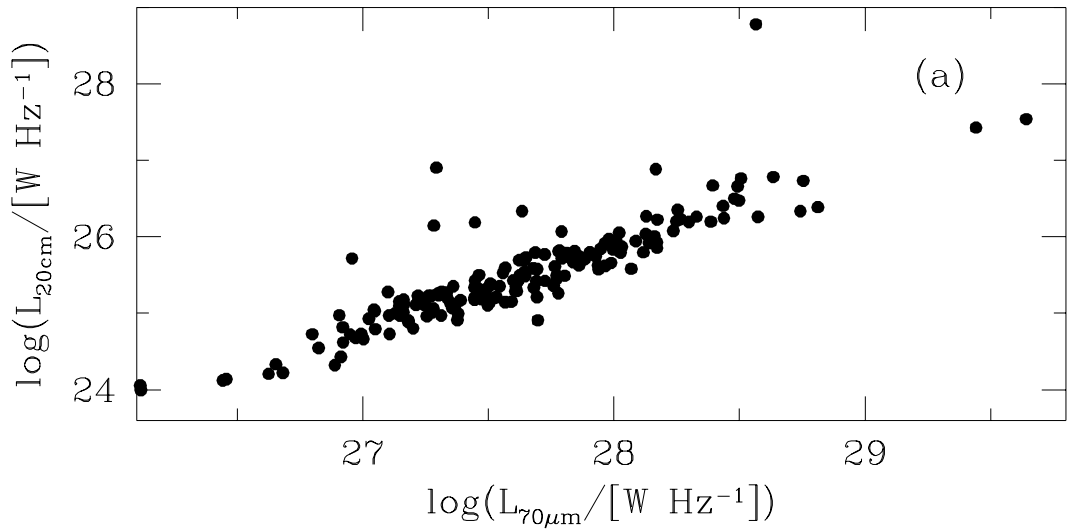
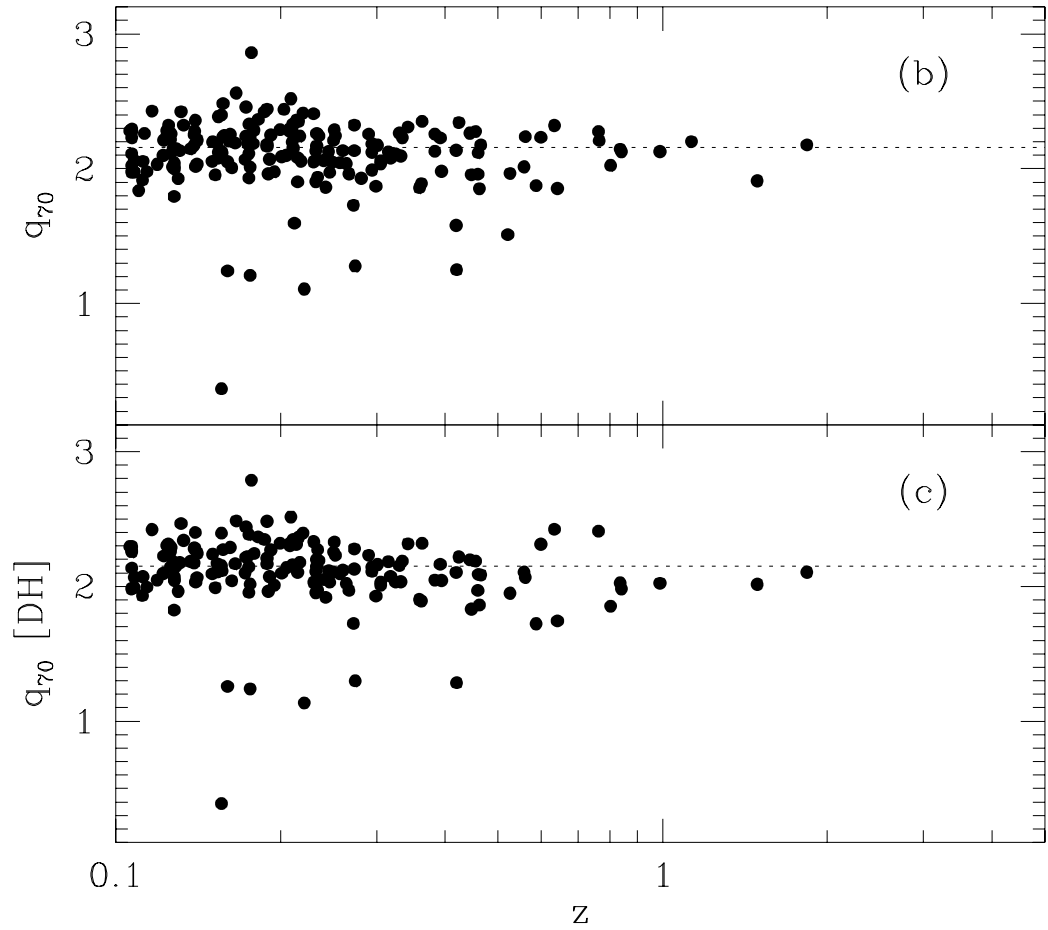
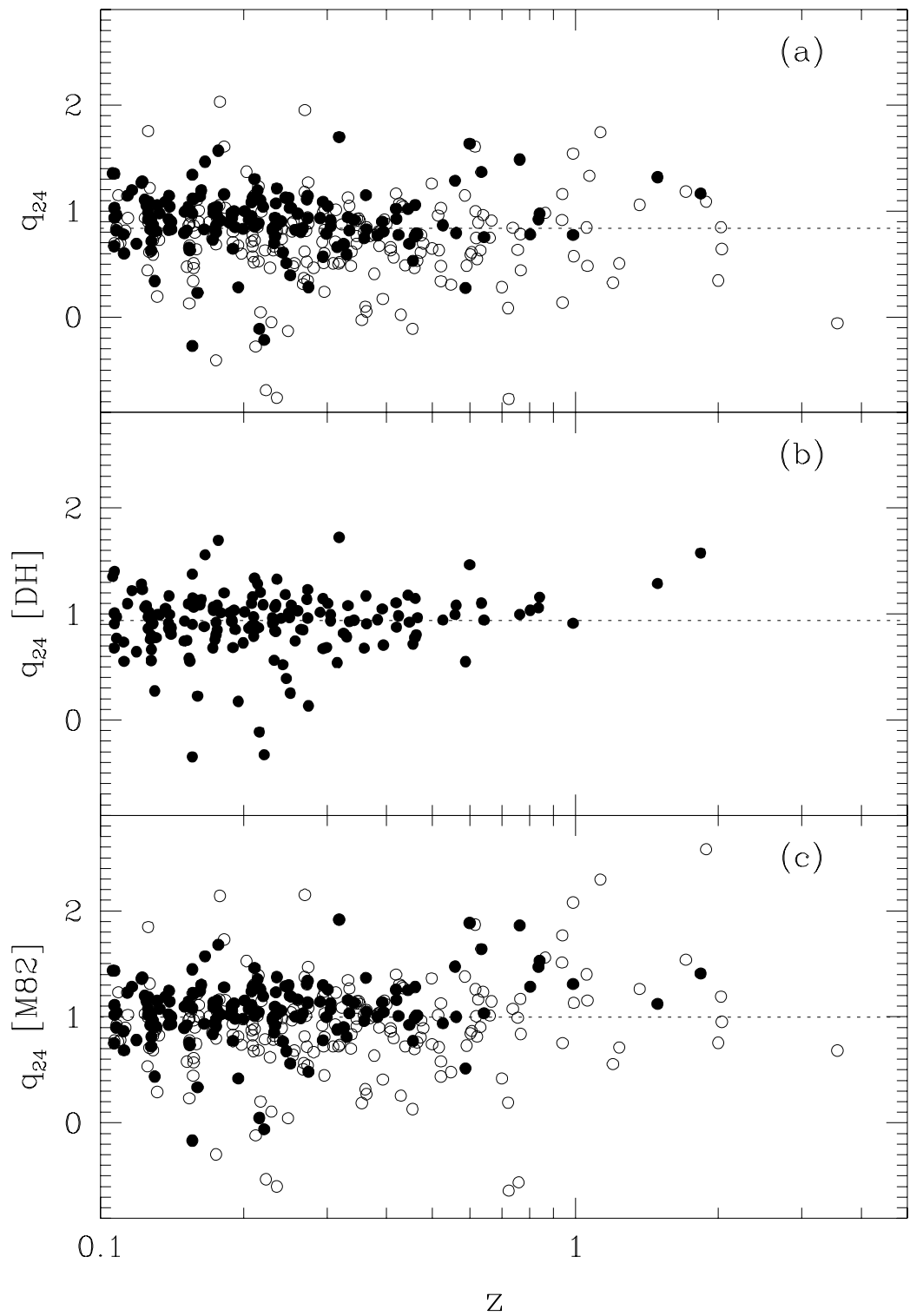
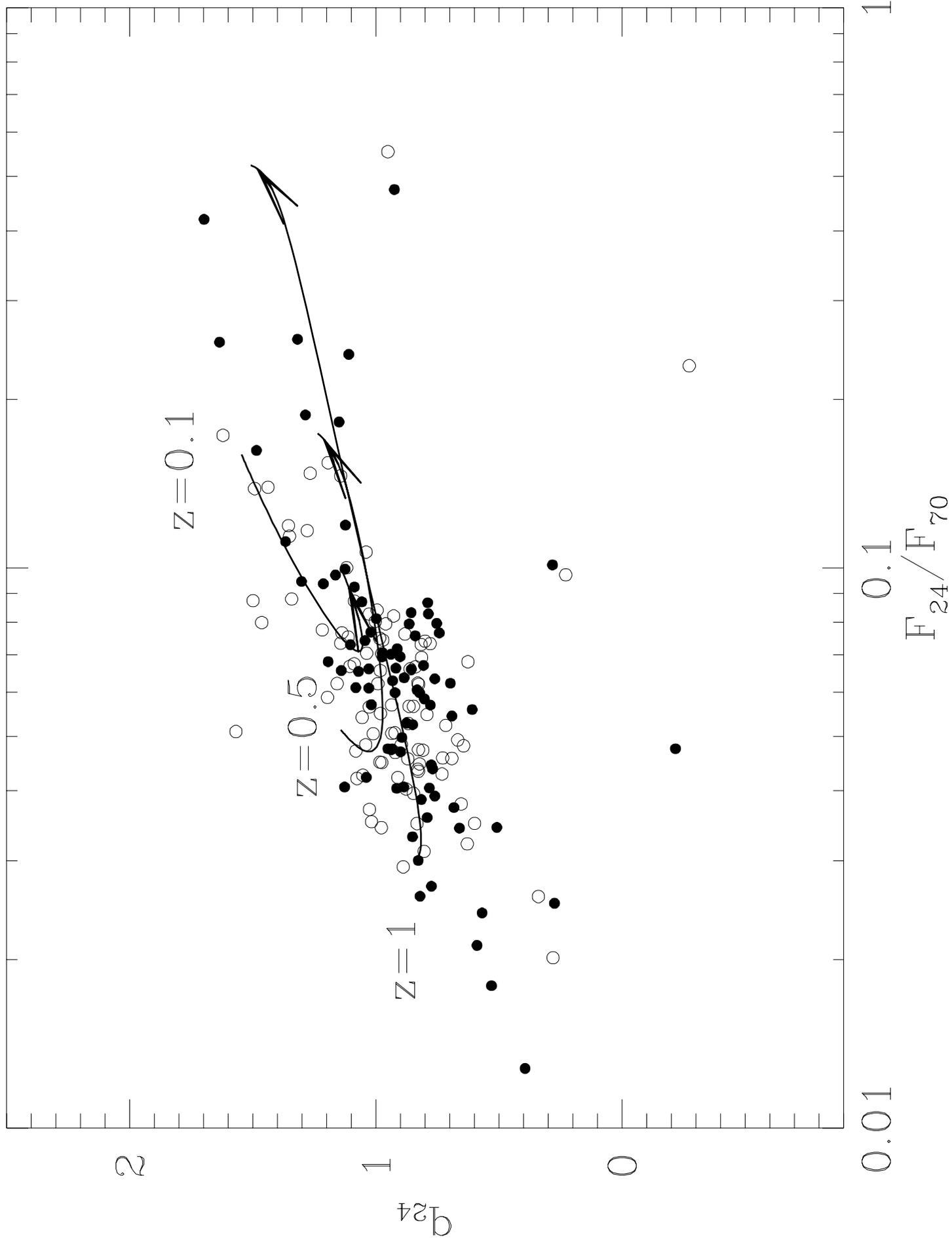
\end{document}